\documentclass[letterpaper,twocolumn,10pt]{article}
\usepackage{usenix,epsfig,endnotes}

\usepackage{booktabs}
\usepackage{multirow}
\usepackage{blindtext}
\usepackage[inline]{enumitem}
\usepackage{xcolor}
\usepackage[ampersand]{easylist}
\usepackage[]{algorithm2e,algorithmicx}
\usepackage{graphicx}
\usepackage{makeidx}

\usepackage{hyperref}

\usepackage[T1]{fontenc}
\usepackage[utf8]{inputenc}

\begin{document}

%don't want date printed
\date{}

%make title bold and 14 pt font (Latex default is non-bold, 16 pt)
\title{\Large \bf
Aurora: Providing Trusted System Services for Enclaves
\\
On an Untrusted System
}

\author{
{\rm Hongliang\ Liang, Mingyu Li, Qiong Zhang, Yue Yu, Lin Jiang, Yixiu Chen}\\
Beijing University of Posts and Telecommunications
% copy the following lines to add more authors
% \and
% {\rm Name}\\
%Name Institution
} % end author

\maketitle

% Use the following at camera-ready time to suppress page numbers.
% Comment it out when you first submit the paper for review.
%\thispagestyle{empty}

\subsection*{Abstract}
Intel SGX provisions shielded executions for security-sensitive computation, but lacks support for trusted system services (TSS), such as clock, network and filesystem. This makes \textit{enclaves} vulnerable to Iago attacks~\cite{DBLP:conf/asplos/CheckowayS13} in the face of a powerful malicious system. To mitigate this problem, we present Aurora, a novel architecture that provides TSSes via a secure channel between enclaves and devices on top of an untrusted system, and implement two types of TSSes, i.e. clock and end-to-end network. We evaluate our solution by porting SQLite and OpenSSL into Aurora, experimental results show that SQLite benefits from a \textit{microsecond} accuracy trusted clock and OpenSSL gains end-to-end secure network with about 1ms overhead.

\section{Introduction}
%\textbf{Big Picture}
Cloud environments play an important role in pervasive computing. Cloud security has been a serious concern. Secure cloud computation is the problem of uploading data and code on remote servers owned and maintained by an untrusted service provider. To overcome this threat, Intel provisions Software Guard Extensions (SGX) technology that establishes a secure container that protects the integrity and confidentiality of desired computation. SGX enforces strong isolation in memory for security-sensitive compartments in an application, called \textit{enclaves}, from untrusted privileged system. Many protection architectures~\cite{DBLP:journals/tocs/BaumannPH15,schuster_vc3:_2015,DBLP:conf/osdi/HuntZXPW16,DBLP:conf/osdi/ArnautovTGKMPLM16,DBLP:conf/icdcs/NguyenG17,7545819} utilizing SGX have been emerging in the context of multi-tenant cloud environments.

%\textbf{Challenges}
An enclave has no direct access to any hardware or OS resources, such as network, storage and clock. In order to gain access to external resources, an enclave must exit to untrusted code and depend upon the untrusted system through system call interfaces, including sockets, file I/O and memory mapped I/O. Unfortunately, this programming model does not provide any secure Input/Output measurements according to the Intel Software Developer's Manual~\cite{Corporation2016Intel}. This unreliable dependency exposes a large attack surface known as Iago attacks, such as system call snooping and I/O traffic analysis.

%\textbf{Motivation}
Due to the intrinsic limitations of SGX, existing SGX-based projects depend on the untrusted OS services, such as time service and network service. State-of-the-art in-enclave library OSes including Haven~\cite{DBLP:journals/tocs/BaumannPH15}, Graphene-SGX~\cite{DBLP:conf/usenix/TsaiPV17} and Panoply\cite{shinde_panoply:_2017} use the untrusted system clock. However, they cannot detect timestamp forgery attack. Blockchain systems such as Town Crier~\cite{DBLP:conf/ccs/ZhangCCJS16} depend on an external relay to obtain a trusted time, but the latency is uncertain due to complex network environments. Intel offers a second-accuracy trusted clock for enclaves, but the time value is course-grained and not absolute~\cite{intel_timer}. Instead, Intel SGX SSL~\cite{intel_sgx_ssl} uses the untrusted \textit{ftime} to achieve millisecond-accuracy time. Other cryptography libraries like TaLos~\cite{TaLoS}, mbedTLS-SGX~\cite{mbedtls} intended for enclave applications also face the same problem.

Likewise, secure network TSSes are also important. One approach is that Haven and Graphene-SGX use a network shielded module to verify results returned from the untrusted host, which is non-trivial and error-prone because of the complexity of network protocol stack~\cite{DBLP:conf/acsac/Bellovin04}. Furthermore, a compromised network stack is susceptible to traffic injection and communication inspection. Another approach targeting network privacy such as SGX-Tor is also built upon the untrusted network I/O interfaces, allowing untrusted systems to compromise the anonymity of Tor’ hidden service by endpoint traffic analysis ~\cite{DBLP:conf/uss/SunEVLRCM15, DBLP:conf/ccs/MittalKJCB11}.

This situation urges us to contemplate the following question: \textbf{\textit{How to provide trusted system services for enclaves on an untrusted system, in order to build safer shielded executions?}} This question is fairly important, since almost all SGX-based projects exclude the underlying system from trusted computing base (TCB) while depending on it for system services. This paper does not cover all system services. We concentrate on \textit{accurate absolute} time and \textit{end-to-end} secure network, which are essentially significant yet rarely discussed in current SGX-based trusted computing ecosystem.

%\textbf{Solution}
In this paper, we describe \textbf{Aurora}, a novel framework that safeguards TSSes for enclaves. Aurora is an architecture that secures messages from devices to enclaves and vice versa. Aurora leverages system management RAM (SMRAM) as a privileged \textit{"enclave"}, to directly interact with devices without involvement of the untrusted system. The SMRAM is tamper-proof to any other software due to the architectural hardware protection. Aurora delegates security-critical services to system management interrupt (SMI) handler located in SMRAM. The SMI handler can safely access devices while keeping transparent to the rest of the system.

Both system management mode (SMM) and SGX provide hardware-based strongly isolated memory region. In fact, the SGX feature must be enabled by BIOS from OEM vendors and setup properly by BIOS firmware to configure the desired size of enclave page cache (EPC). We hold the insight that BIOS is the key to the availability of SGX and can offer a competent SMI handler to provide TSSes for enclaves in a secure manner.

Aurora supports TSSes for enclaves with the following properties:
\begin{itemize}%[noitemsep]
\item \textbf{Extensibility:} Aurora is designed to be a general TSS framework to offer a broad range of devices with the ease of driver portability. Currently Aurora supports five hardware timer and a commodity network adapter for enclave programs.
\item \textbf{Security:} Aurora utilizes SGX EPC memory and SMRAM to provide strong confidentiality and integrity. Aurora creates a secure session between the SMI handler and SGX enclaves. It prevents information leakage by leveraging identity authentication, hardware encryption scheme and constant-time protocol.
\item \textbf{Transparency:} Aurora is adversary-unaware and deployment-friendly. It does not require any modification on commercial hardware and operating systems. It supports POSIX APIs for enclave programs to request TSSes.
\item \textbf{Efficiency:} Aurora makes several optimization to minimize run-time overhead on I/O requests compared with monolithic kernel. It overcomes the architectural limitations of Intel SGX by introducing event-based notification, exit-less interrupts and batching SMI calls.
\end{itemize}

To summarize, we make the following contributions:
\begin{itemize}%[noitemsep]
%\item[$\bullet$] We make an exploration and comparison on possible approaches in terms of how to provide TSSes for enclaves on untrusted systems (\autoref{exploration}).
\item[$\bullet$] We present \textit{Aurora}, a new architecture to provide TSSes for enclaves, with the benefit of extensibility, security, transparency and efficiency (\autoref{arch}).
\item[$\bullet$] We design and implement a high-precision and attack-aware time TSS (\autoref{time service}) and an end-to-end network TSS (\autoref{network_service}) based on Aurora. To the best of our knowledge, we are the first to support such TSSes for enclaves.
\item[$\bullet$] We evaluate Aurora using real-world applications: SQLite and OpenSSL. (\autoref{evaluation}).
\end{itemize}

% 2. Background
\section{Background }
To provide system services for enclaves, Inktag~\cite{DBLP:conf/asplos/HofmannKDLW13} and Sego~\cite{DBLP:conf/asplos/KwonDLHXW16} utilized a trusted HV to ensure the integrity of files for high-assurance processes (HAPs). A hypervisor (HV) that has higher privilege than the untrusted system can regulate the guest and therefore provide TSSes for applications running on the guest. The major benefit of this solution is that a trusted HV can use virtulization to isolate memory region of untrusted guests and the input–output memory management unit (IOMMU) to circumvent direct memory access (DMA) attacks from malicious devices. 
Although an HV's code base (e.g. Xen~\cite{DBLP:conf/sosp/BarhamDFHHHN03}) is usually smaller than a commodity kernel, there are many security vulnerabilities over the years~\cite{Xen}, which makes them as attractive attack targets. Furthermore, these studies did not address the clock and network TSSes on untrusted systems.

\subsection{System Management Mode}
System management mode (SMM) is the most privilege mode available in all x86 platforms. The CPU enters SMM upon a system management interrupt (SMI) and executes the system management handler (SMI Handler), a special segment of code loaded from the BIOS firmware into system management RAM (SMRAM). An \textit{RSM} (resume) instruction is executed at the end of the SMI hander to switch back to Protected Mode. The OS or HV is essentially suspended while the SMI handler executes. This isolated execution provides transparency to the operating system. The SMI handler can run any instruction and has complete access to all devices and control over interrupts regardless of any protections established by the OS or HV. We expand the SMI Handler to design Aurora's supervior \autoref{arch}.

The critical code and data inside SMRAM is inaccessible by other modes. The SMI handler requires only a small trusted code base and is safe from corruption after booting when properly configured.

SMM-based protections~\cite{DBLP:conf/esorics/ZhangWLS14} outperform the kernel-based and HV-based approaches with its overwhelming transparency. To the best our knowledge, researchers have not use SMM to supply TSSes for TEEs.

\subsection{Software Guard Extension}
Intel’s SGX provides trusted execution environments (TEE) called enclaves. Enclave code and data reside in specialized protected memory called enclave page cache (EPC). Enclave code can access the memory outside the enclave. As enclave code is only allowed to be executed in user mode, any interaction with the device must execute outside of the enclave and through untrusted system calls. SGX enables a threat model where one only trusts the Intel CPUs and the code running inside the enclave(s).

Previous studies use library OSes~\cite{DBLP:journals/tocs/BaumannPH15, DBLP:conf/usenix/TsaiPV17, shinde_panoply:_2017} to provides TSSes such as thread scheduling. For I/O services such as network, the library OSes have to forward such requests from applications to the untrusted host.

\subsection{Threat Model}
At the hardware level, we assume that processor is implemented correctly and equipped with flawless SGX functionality. We assume that the SMI handler to be loaded into SMRAM can be only initialized at boot time by BIOS firmware from trusted OEM vendor and be made inaccessible from other operating modes. The SMRAM is prevented from unauthorized memory accesses (e.g. cache poisoning~\cite{itl_smm_cache_attack}). The devices are trusted and can operate correctly when configured properly. Hardware attacks on devices are not considered.

At the software level, we consider a powerful adversary that controls the entire system software stack. The adversary can arbitrarily manipulate the OS, including scheduling desired threads, simulating signals and software exceptions. He may use any I/O commands to trigger hardware interrupts and issue DMA write requests at will. Denial of service attack (DoS) is out of the scope, where the untrusted system is able to disable SMI.

Our trusted computing base (TCB) does not consist of entire BIOS, only the SMI Handler inside SMRAM is trusted. While we do not claim to prevent all covert channel attacks, Aurora does mitigate cache attacks on cryptographic process and timing attacks on requests for TSSes. We assume that Intel SGX SDK and toolchains are from the trusted source.
We do not consider the enclave linked against Aurora's library to be bug-free~\cite{DBLP:conf/uss/LeeJJKCCKPK17}, therefore we consider potentially compromised enclaves. Enclaves that intend to leak secrecy is not considered.

%\subsection{Alternative Design}\label{exploration}
%In this section, we try to seek a suitable architecture and ask the question: \textbf{\textit{Can previous studies on protecting trusted executions environments (TEE) from untrusted systems be applied to provide TSSes for enclaves too?}}

%But virtulization-based approaches cannot defend against BIOS malware~\cite{}.

%\textbf{SGX-based Approach}
%For studies that supports network TSS using TLS encryption inside enclaves such as SGX-Tor~\cite{DBLP:conf/nsdi/KimHHKH17} and SCONE~\cite{DBLP:conf/osdi/ArnautovTGKMPLM16}, send encrypted payloads with untrusted sockets. This method cannot protect the payload from traffic analysis or injection at the endpoint because network stack is controlled by the untrusted system.
%An better solution for network TSS is to port a user-level stack into enclaves to protect stack itself as well. However, the network driver is also left unprotected. Finally, we turn to user-space driver solutions such as \textit{vfio} and \textit{uio}, but the interrupts from devices are still handled inside kernel and a malicious kernel is able to launch interrupt spoofing attack~\cite{zhou_building_2012} to compromise network TSSes.

%\textbf{SMM-baesd Approach}

%\textbf{Discussion.}

\section{Aurora Design}
\begin{figure}[t]
\centering
\includegraphics[width=0.48\textwidth]{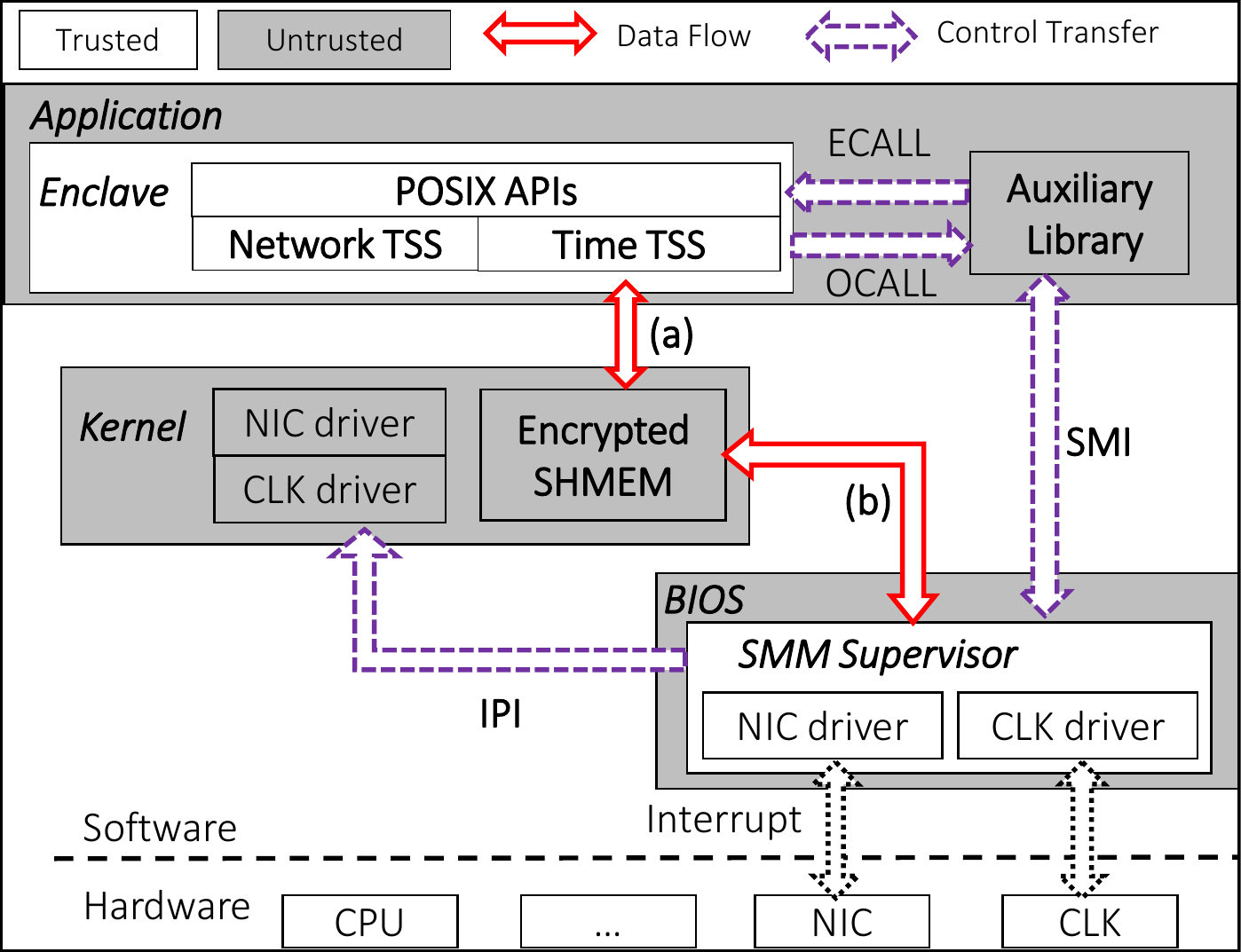} %height=0.25\textheight,width=0.78\textwidth
\caption{The architecture of Aurora}
\label{fig:arch}
\end{figure}

In order to provide reliable TSSes, we use the combination of SGX and SMM because of its hardware enforced protection. Since the protection is guaranteed at the opposite ends of Intel x86's privilege-level model, we name our framework after \textit{\textbf{Aurora}}, which takes place in the polar region. Aurora takes advantage of architectural support to build TSSes for enclaves. Aurora is composed of three components: the SMM Supervisor (SSV), in-enclave TSS library and a secure channel established between the SSV and enclaves. Figure \ref{fig:arch} depicts the overall architecture of Aurora.

Aurora adopts the philosophy of \textit{delegate and emulate} to support TSSes. The process of one TSS request can be briefly described in the following three steps. First, Aurora builds a secure channel between SGX enclaves and SMM Supervisor (SSV). Second, it allows enclaves to send TSS requests to SSV using a secure session. Last, SSV executes the desired operations to repsonse the TSS. This is conceptually similar to a normal system call or VM hypercall. We name it a SMM call.

\subsection{Architecture Overview}\label{arch}
%In order to provide reliable TSSes, both drivers and related software stack need be carefully protected. The SMM can benefit us with an SMRAM that isolates its SMI handler from any privilege software, therefore it can prevent the drivers for TSSes from corruption. The SGX protection offers user-level tamper-proof protection from the untrusted system, which can accommodate the data processing logic. The collaboration of SMM and SGX can be used to provide TSSes with strong security guarantees for its architectural isolation and protection.

\begin{enumerate} [label=(\arabic*)]
\item The SMM Supervisor (SSV) can take control of interrupts from the untrusted system and it contains lightweight drivers that handle device interrupts. The SSV is responsible for determining the source of interrupts and dispatching interrupts to the untrusted system via an IPI or the destined enclave. In order to provide OS-transparent TSSes, a carefully designed driver specification is proposed (\autoref{driver}).%The SSV exchanges messages with enclaves via encrypted byte streams using shared memory.
\item The Aurora TSS library supports standard POSIX APIs (such as BSD sockets) for enclaves. This complements unsupported APIs of Intel SGX SDK, helping developers to port vanilla applications into SGX programming model.
\item The secure channel prevents the TSS secrets leakage from the untrusted system. To achieve so, Aurora introduces a shared memory and implements a FIFO module to exchange encrypted messages between SSV and enclaves. When the user inside the enclave requests a TSS, it will leave the enclave via an \textsl{OCALL} and invoke the auxiliary library, which then \textit{"calls"} the SSV via an SMI.
%The session protects the man-in-the-kernel attack that fakes its identity and tries to infer secrets by observing. To minimize the overhead, Aurora implements batched mechanism using two FIFOs and an auxiliary library that serves to forward interrupts when necessary. We decouple the existing software stack from the complex kernel. That is, we lift data processing logic up into enclaves and sink device drivers down to SMM.
\end{enumerate}

\textbf{Aurora's Workflow.}\label{workflow}
We take the time TSS as an example to illustrate Aurora's workflow. We break down the process into the following phrases:
\begin{enumerate*}[label=\arabic*)]
\item The enclave starts to request the time TSS and this message is encrypted inside EPC;
\item The enclave’s FIFO module copies the encrypted message into the shared memory;
\item The auxiliary library triggers an SMI and switches the system to SMM;
\item The SSV’s FIFO module copies the encrypted request to SMRAM;
\item The encrypted request is successfully decrypted and the SSV invokes the clock driver;
\item The clock driver reads the real-time value from the hardware and assembles the data as a response message inside SMRAM;
\item The response information is encrypted inside SMRAM and ready to be sent back;
\item The SSV’s FIFO copies the encrypted message into the shared memory;
\item The SSV clears the secrets and execute the RSM instruction to return from the SMM;
\item The system is resumed and enclave’s FIFO module copies the encrypted response into the EPC;
\item The response is decrypted inside EPC and the enclave program successfully obtain the trusted time value.
\end{enumerate*}

\subsection{SMM Supervisor (SSV)}
The SSV is the fundamental component in Aurora. When SSV receives a TSS request, it takes control of the target device, including interrupts, to prevent untrusted systems from interrupt spoofing attacks. Since it may take charge of more than one device, it needs to determine which device is to be serviced and invoke the matching driver (\autoref{driver}).

\textbf{Interrupts Interception.}
SSV intercepts device events by configuring the \textit{Redirection Table} defined in I/O Advanced Programmable Interrupt Controller~\cite{Intel_82093AA_IO_APIC}. This results in that selected device interrupts are firstly routed to the SSV. Such techniques are also applied in~\cite{DBLP:conf/securecomm/EmbletonSZ08,zhang_trustlogin:_2015}. When the SMI is triggered, SSV distinguishes the interrupt source and determines which domain (the system or an enclave) to be notified. It then forwards the interrupt using Inter-processor interrupts (IPIs).

\textbf{Device Addressing.}
In order to perform the operation, the driver invoked by SSV needs to know the exact I/O ports of the target device. At system boot, the BIOS firmware collects device information and performs configuration transactions for peripherals it detects. Meanwhile, the SSV records their memory mapped I/O (MMIO) base addresses, thereafter the corresponding driver can manipulate the mapped configuration space.

\textbf{Boundary Checking.}
During run-time, the driver may require more memory to finish its purpose. SSV supports a sanity-checking dynamic heap to avoid secret leakage. SSV checks the memory boundary that the driver accesses to ensure the driver dose not leak information outside of SMRAM.

\subsection{SMM Driver}\label{driver}
Because the kernel drivers are coupled with the untrusted system and lack effective isolation from the rest of the privilege software, we cannot reuse them.
We port commodity drivers with minimal necessary functionalities and protect them inside SMRAM. We see such effort feasible for three reasons. First, SMM mode is similar to kernel mode where privileged CPU instructions are available. Second, the mechanism of interrupt rerouting helps SMM driver design to concentrate on the interrupt handling rather than device initialization or resource management. Third, previous works have shown how to extract drivers from commodity OSes, including user space driver implementation~\cite{DBLP:conf/usenix/Boyd-WickizerZ10,DBLP:conf/sosp/SwiftBL03} and driver isolation~\cite{DBLP:conf/asplos/GanapathyRBSJ08,richter_isolating_2016}. In order to cooperate with Auroa's SSV, Aurora requires the drivers to obey a specification as follows.

\textbf{Driver Specification.} There are three APIs that a driver needs to provide in order to accomplish the TSS for enclaves: \textit{aurora\_probe()} makes sure the device is available and ready to be used, \textit{aurora\_write()} and \textit{aurora\_read()} are used to process data flow from and to the enclaves. The read and write operation logic can be simply extracted from the interrupt routines in commodity drivers. We do not allow the driver to re-initialize the device or expose richer interfaces to enclaves, because these \textit{IOCTLs} may be abused by a malicious enclave to interfere the operations of the normal system.

\textbf{Guideline.} In order to achieve transparency to untrusted systems and avoid unexpected misuse of critical resources, we enforce such confinements on the SMM driver design:
\\
\textit{Pre-condition:} Suspend OS. Save context. Enter SMM.
\begin{enumerate}[label=<\arabic*>]
\item If only a read to the device is required, jump to <3>.
\item Save the device context, including control registers.
\item Perform operations (read or write). Trigger interrupts immediately to finish the commands.
\item if device context is saved, restore it.
\end{enumerate}
\textit{Post-condition:} Exit SMM. Restore context. Resume OS.

\subsection{Secure Session}
During a TSS request, we treat the SSV as the server and the enclave as the client.
To provide reliable TSSes for enclaves, we design a secure session that protects the messages between the SSV and enclaves. In this section, we describe the the lifecycle of a secure session.

\subsubsection{Establishment}
The initial stage of the secure session is to make sure whether both parties involved in the session are trusted or not. Since the untrusted system is located between the two parties, it can launch the man-in-the kernel attack. We introduce identity authentication to avoid this. When both parties are acknowledged, they exchange the symmetric secret key using key agreement.

\textbf{Enclave Identity Authentication.}
An enclave can issue an SMM call using port writing to request SSV's service. The privileged software also has this capability and may launch SMI-specific fuzzing attacks on the SSV. To prevent abuse of this mechanism, Aurora uses SGX credentials during remote attestation to verify the enclave's identity. As a result, only the legitimate enclaves have the right to establish a secure channel with the SSV. In this sense, requests from untrusted privileged programs and unauthorized enclaves will be discarded.

\textbf{Supervisor Identity Authentication.}
The privileged system can also fake the identity as an SSV since it can emulate an SMI and try to \textit{handshake} with the intended enclave. We design an approach similar to enclave's remote attestation to verify SSV's genuine identity. During startup, the BIOS communicates with Intel remote server to establish a secure channel using public key infrastructure (PKI). The BIOS computes a token on the image (SSV) to be loaded into SMRAM and sends its hash signature to verify its integrity and prove its identity.

The remote attestation is significantly vital to Aurora's key agreement phrase because it makes sure that:\\
\begin{enumerate*}[noitemsep]
\item Aurora is running on real SGX-enabled hardware;
\item the SSV is neither counterfeit nor outdated;\\
\item the integrity of SSV’s initial state is not compromised.
\end{enumerate*}

\textbf{Key Agreement.}
When an enclave is securely launched and proves its identity through remote attestation, the remote server plays the role of certificate authority. The enclave generates a string of random bytes as the symmetric secret key. The remote server exchanges this key between the SSV and the enclave instance on the same machine. At this point, a unique session has been successfully established. We then use AES-GCM encryption scheme for further message exchange. The enclave can reset the channel at any time to mitigate possible covert-channel attacks.

\subsubsection{Communication}
Aurora's objective is to provide TSSes with the properties of security and efficiency. To achieve security, Aurora prevents the untrusted system from inferring secrets by introducing \textit{data obfuscation} and \textit{hardware encryption}. For efficiency, Aurora adopts \textit{FIFO mechanism} to eliminate the expensive transitions.

\textbf{Data obfuscation.}
If an enclave requests the same TSS twice and the results happen to be the same, the encrypted messages will also be identical if the session key is not changed. A malicious attacker can infer secrets by observing such covert channels. To prevent such secrecy leakage, Aurora pads/reassembles all messages to the same length. Currently we set the fixed size to 4KB, same as an EPC page size. The message buffers must start with page-aligned address. This restricts possible information leakage at coarse-grain page-level. This property is also referred to as data oblivious~\cite{Ohrimenko2016ObliviousMM}.

\textbf{Hardware encryption.}
The implementation of encryption is an attractive target for attackers~\cite{DBLP:conf/acns/BiryukovDC17}. In fact, Intel has extended x86 ISA with Advanced Encryption Standard New Instructions (AES-NI) Set and claims to prevent known side-channel attacks. We make use of it to address this concern and use constant-time implementation to defeat cache timing attacks.

\textbf{Shared Memory.}
The SGX memory model is asymmetric: the enclave code can access the entire process address space. As a result, Aurora library can safely compute on the shared memory without exiting enclave mode. On Linux we use a kernel module that allocates contiguous physical memory (currently 1MB) and exports an interface as a character device \textit{/dev/aurora}. The auxiliary library will \textit{open} and \textit{mmap} this device to its address space. If the untrusted system launched memory-based Iago attacks such as mapping a fake device, we treat it as a DoS attack and the secure session will not be setup.

\textbf{Protocol Interface.}
Aurora provides an uniform abstraction API: \textit{IOCTL(DEVICE, EPID, OPERATION, PAYLOAD)}. The first argument indicates the device that an enclave desires to communicate with. The second identifies the authentication of requestor. Intel Enhanced Privacy ID (EPID) is used for signature verification; we use it to distinguish enclaves. The third stands for the operation type to be performed, it can be either of the three: PROBE, READ and WRITE. The reason is explained in~\autoref{driver}. The last argument is the message data to be attached. All the arguments will be marshalled and encrypted into the shared memory and can be only understood by the SSV.

%\begin{figure}[t]
% \centering
% \includegraphics[height=0.25\textheight,width=0.45\textwidth]{FIFO.png}
% \caption{Aurora's FIFO mechanism}
%\end{figure}
\textbf{FIFO module.}
For TSSes such as network, the driver may receive more than one Ethernet frames upon one interrupt. To maximize throughput, Aurora supports a buffering mechanism using two dedicated, lock-free FIFOs. This is also useful in the case of high frequent transitions when combined with asynchronous, batched mechanism. The user can set a desired threshold, so the auxiliary library will coalesce the interrupts and reduce the times of costly context switches. For instant requests such as time TSS, the requestor can directly issue an SMM call which skips the FIFO buffering. Such optimization mechanism is also applied in ~\cite{DBLP:conf/osdi/ArnautovTGKMPLM16,weisse_regaining_2017}.
%To eliminate the costly context switches between SMM and enclave mode, Aurora supports an asynchronous, batched interface. This interface consists of two dedicated, lock-free FIFOs, as depicted in figure 2. This is especially useful in the case of high frequent transitions. The user can set a desired threshold, so the auxiliary library will coalesce the interrupts and reduce the times of context switches. For instant requests such as time services, the requestor can directly issue an SMM call which skips the FIFO buffering. This optimization mechanism is also applied in ~\cite{DBLP:conf/osdi/ArnautovTGKMPLM16,weisse_regaining_2017}.

\subsubsection{Teardown}
When the requestor enclave process exits, the SSV will terminate the session and release resources. It will disable its service when no enclave is requesting services, thus making no impact on the system.

\subsection{Time TSS}\label{time service}
%\textbf{Time service is important!}
Clocks and timers are simple yet important devices. The dysfunction of a timer can compromise the normal application's logic, such as connection expiration and certificate revocation. With a trusted clock source, one is able to create trusted timestamps. This is especially significant in financial transaction (e.g. blockchain) and secure network (e.g. TLS authentication). Therefore, we extend our framework to provide absolute, high-precision and attack-aware time TSS.

\textbf{Absolute and High Precision.}
We use {Real-Time Clock (RTC)} to provide the absolute wall-clock time for enclaves. If the RTC clock is not calibrated, it falls back to offer a reference clock, which is the same functionality of Intel's \textit{sgx\_get\_trusted\_time()}, but provide time with a much smaller delay, as shown in~\autoref{sqlite}. When higher precision is required, we use the High Precision Event Timer (HPET) to compute the \textbf{tv\_usec} value in \textit{timeval} structure.

\textbf{Multiple Sources.}
We also use other hardware clock sources including Programmable Interval Timer (PIT) and Advanced Programmable Interrupt Controller (APIC) timer to support Aurora's Time TSS. We reference Time Stamp Counter (TSC) to estimate the latency of other timers and adjust their value. We provide all timer values to enclaves.

\textbf{Validation.} The hardware timers and clock can be controlled by the malicious OS. The diversity of timers offers the ability for enclaves to validate the credibility of time value. That is, when any of the time values violates the monotonic rule, we assume a time attack.

In a word, Aurora not only provides the ability to request time directly from the hardware, but also gains capability to verify its authenticity.

\subsection{Network TSS}\label{network_service}
%\textbf{Network stack needs to be protected too!}
We employ SMRAM to protect the integrity of the NIC driver. However, in order for an enclave to communicate over the network, a protocol stack is also required and must be protected as well, making an end-to-end network TSS feasible. Previous studies~\cite{DBLP:conf/nsdi/JeongWJJIHP14,DBLP:conf/apnet/HuangGLWLLL17,DBLP:conf/sosp/EickenBBV95,DBLP:conf/sigcomm/MarinosWH14,DBLP:conf/sosp/PrekasKB17} has proved that adopting kernel-bypass network has advantages in both scalability and performance, but this approach suffers from lack of protection for network stack when running on an untrusted system. We retrofit their design by protecting the network protocol stack using SGX isolation. We connect the data link layer in SMM with the network layer inside enclaves using Aurora’s secure session. We reason about several design choices on Aurora's network TSS.

\textbf{Per-Thread Stack:} SGX allows multiple threads to reenter the enclave mode, thus an in-enclave stack can serve many threads simultaneously. This requires synchronization mechanism in the application layer. We do not choose this design because mutual exclusions and conditional variables from Intel SGX SDK are built upon untrusted system service. The untrusted OS is responsible for scheduling all threads and can deliberately construct race conditions, leading to preemption-based attack~\cite{DBLP:conf/esorics/WeichbrodtKPK16} such as Time-of-Check-to-Time-of-Use attack.

Instead, we make every thread have its own copy of stack to avoid critical section. It also guarantees that a malicious or misbehaving enclave thread can only hurt itself without corrupting concurrent threads. Likewise, Aurora uses only one thread in stack layer to process packets. This means all networking activities are synchronized. The alternative design that one master stack thread and multiple slave application threads is not considered because currently EPCs cannot be shared between enclaves and using system Inter-Process Communication (IPC) is unsafe and introduces more overhead. To summarize, our design ensures correct behavior and simplifies developing effort.

\textbf{Multistack Coexistence:} It is possible for users to spawn multiple enclave instances to request network service at the same time. We use MultiStack~\cite{DBLP:journals/ccr/HondaHRAR14} approach to address this issue. Each enclave stack has its own ring buffer, which accommodates encrypted Ethernet frames through the secure channel from SSV.

\textbf{Flow Multiplex.} The SSV is responsible to multiplex corresponding incoming traffics flow for enclaves. There are several approaches that the SSV can use:
\begin{enumerate*}[label={\roman*)}]
\item \textit{Multi-MAC Addresses.} The network stack can share the same IP address with the host, but the port numbers may conflict. In order to assign an IP address to each individual stack, SSV turns the NIC device into promiscuous mode and generates unique MAC address to obtain a public IP through DHCP protocol. However, this brings more workloads to NIC hardware.
\item \textit{Network Address Translation (NAT).} NAT allows to hide enclave networks from outside and is fit to implement virtual private network (VPN). In this case, the SMM network driver acts as a middlebox.
\item \textit{Packet Introspection.} Both IPv4 and IPv6 reserve an \textit{IP\_OPTIONS} field, we can use this as a distinct flag to distinguish the flows. We choose the third because it is flexible and easy to deploy.% and can be coordinated with other security mechanisms for enclave-level access control in the future. See our discussion in ~\autoref{discussion}.
\end{enumerate*}

\section{Implementation}

\subsection{SMM Supervisor}
We modified the SMI handler of SeaBIOS 1.10.0 and implemented our SSV.

\textbf{Hardware Encryption.} In order for AES-NI instructions to be executed within SMM, we enable the SSE bit in XCR0 register when switching to the SMM mode. With the assistance of SSE and AES-NI, we boost encryption performance and effectively reduce the response time of SSV. On our hardware, the AES-NI based 256-GCM encryption and decryption on 4KB data in SMM mode costs 9us on average while the non-AES-NI 128-GCM implementation of Intel IPP Crypto library~\cite{Intel_IPP_2017_Update_2} from Linux SGX SDK costs 597us.

\textbf{Secure Memory Manager.} We implemented a secure heap manager for the purpose of dynamic memory allocation. The manager makes sure that the memory assigned to the driver must be located within SMRAM. The drivers is responsible to free all allocated memory upon completion. To avoid possible memory leaks, each time before returning to the protected mode, the manager will free all chunks of blocks.

\subsection{Time TSS}
We reused timer and clock driver implementations of SeaBIOS, including RTC, HPET, PIT, TSC and APIC Timer. To minimize the preemption time of time TSS in SMM, Our driver is only responsible for obtaining the raw data from hardware clock and timers and we implement time processing logic in enclave library. The time TSS library bookkeeps the time values in order to detect the time attack.

\subsection{Network TSS}

\textbf{Network Device Driver.}
%In our current prototype, Aurora used AMD PCNET, a PCI-based network adapter. The PCNET NIC supports direct memory access (DMA) and ring buffers to efficiently exchange frames with the driver. Since the SMRAM cannot be accessed by the NIC's DMA, our driver has to copy plaintext frames into unprotected memory. To avoid secrecy leakage, When the transmission is done, the driver frees the memory and the heap manager clears this memory to avoid potential leakage.
%\textbf{Networking APIs.}
Currently we used AMD PCNET, a PCI-based network adapter to implement network TSS. Based on SMM driver specification~\autoref{driver},  three APIs in our prototype are implemented as follows:

\textsl{aurora\_net\_probe()} obtains access to the base address register of NIC and indexes TX and RX descriptor rings.

\textsl{aurora\_net\_read()} iterates every single RX ring buffer. Once it finds a packet with a particular pattern in the \textit{IP\_OPTIONS} field, it \textbf{moves} the packet into FIFO module.

\textsl{aurora\_net\_write()} fills next available descriptor with the frame from the source enclave. Unfortunately, the SMM driver does not know which ring is ready, so we suspend the NIC in order to obtain the ring counter from device control registers directly. On its completion, we trigger an immediate send signal to NIC and restore the original values to corresponding registers. %This follows the guideline we proposed in ~\autoref{guideline}.

\textbf{Userland Network Stack.}
We ported lwIP 2.0.3 into the enclave. We chose it because of its modularity, maturity and its small code base. We eliminated any dependence of lwIP stack on operating systems. Our network implementation does not require any \textsl{OCALL} except triggering SMIs. The global pool for holding packets and connection states is configured at compile time. The network stack requires entropy to generate random port number and timestamps to check connection expiration. We obtain randomness from trustworthy RDRAND instruction and time value from our trusted time service.
%\textbf{Entropy and Timestamps:}
%\textbf{OS Independence.}
%\textbf{Memory Management.}

\textbf{Optimization.}
When there are not many TSS requests, polling will waste precious CPU resources. We adopt an interrupt-based mechanism to notify the enclaves when its relevant TSS is ready. Our current prototype uses a Userspace I/O (\textit{UIO}) device to relay the interrupt from SSV. Note that this is an default optimization for the untrusted system. If the system refuses to accept this module, Aurora will fall back to polling mode, requiring more CPU cycles.

\subsection{Code Base}
%\begin{table}
% \centering
%\begin{tabular}{ |c|c| }
% \hline
% Module & Code Size (kLoC) \\
% \hline
% Cryptography & 1119 \\
% SMM Supervisor & 274 \\
% Clock driver & 203 \\
% NIC driver & 162 \\
% Protocol Stack & 46575 \\
% \hline
%\end{tabular}
% \caption{The TCB of Aurora (header files excluded)}
%\end{table}
%Table 2 shows the overall code base of Aurora's current prototype. As shown in the table, the mini drivers contain much fewer code than commodity drivers, which consist of 2312 LoC for the PCNet network card and 368 Loc for a general RTC device.

Our current prototype of Aurora SSV consists of 2122 lines of code in total. Our network driver only consists of 161LoC. It is much smaller than that in the commodity kernel (which is 2312LoC) because our driver concentrates on the interrupt handling (\autoref{driver}), showing that the efforts to port an existing commodity driver into Aurora is quite intuitive and straightforward.

Since both SMRAM and EPCs are scarce memory resource, we measure the final size of resulting images. Our modification on SeaBIOS only add 3.9\% on its total size (120.2KB V.S. 115.7KB). The enclave.so is 696KB in size and 1MB including minimal runtime stack and heap allocation. This implies that we could have roughly 56 instances of Aurora with network support in parallel without impacting EPC paging performance (The system-wide EPC limit is approximately 93MB). Such is suitable for micro-services with critical safety requirements~\cite{DBLP:journals/ieeesp/Fetzer16}.

\section{Evaluation}\label{evaluation}
\textbf{Experimental setup.} To date, the SMRAM in commodity PCs is carefully protected from modification due to security issue. To validate our framework without breaking hardware protection, we leverage virtualization technology to emulate the SMM functionality. The SMM is emulated by QEMU/KVM and SGX is assured by real hardware. We use Dell Inspiron-5577 laptop with Intel(R) Core(TM) i7-7700HQ CPU @ 2.80GHz running Ubuntu 16.04 LTS and SGX SDK 1.9.100 and SGX driver 0.10. We use the QEMU emulator provided from https://github.com/intel/qemu-sgx. We assign one CPU-core to the QEMU. We use the same CPU model and run the same software stack (including the system and drivers) in the guest VM. The virtualized networking environment for SGX-enabled VMs is setup with Open vSwitch 2.5.2.

\textbf{Methodology.}
We evaluate Aurora by answering the following questions:
\begin{enumerate}[noitemsep]
\item Does Aurora offer security guarantees for TSSes? (\autoref{security})
\item What overhead does Aurora introduce to the system? (\autoref{overhead})
\item What is the quality of time and network TSS? (\autoref{time_bench})
\item What is the performance overhead of running real-world applications on Aurora? (\autoref{sqlite}, \autoref{openssl})
\end{enumerate}

\subsection{Security Analysis}\label{security}
We evaluate Aurora's security by answering the following questions:

\textbf{Can Aurora protect integrity and confidentiality of TSS session messages?} During a TSS, both parties in a secure session are located in hardware enforced isolated memory, i.e., SGX EPCs and SMRAM. Any untrusted privilege code can never access nor modify them. The communication channel is protected using an authenticated encryption algorithm (AES-GCM) based on Intel AES-NI instructions. The secret key and plaintext messages are carefully protected inside SMRAM and enclaves.

\textbf{Can Aurora protect itself from DMA Attacks?} SMRAM is strongly isolated and can only be accessed when CPU is in SMM. A hypothetical DMA attack would not be able to corrupt our SMM driver logic protected in SMRAM. Meanwhile, the CPU rejects any DMA transfer within the Processor Reserved Memory (PRM) region. However, the unprotected shared memory is susceptible to DMA attacks. Intel Trusted Execution Technology (TXT) can be used to mitigate this. Instead, we assure data cached in unprotected memory to be always encrypted. Note that the architectural support cannot provide seamless switch between enclave and SMM mode, there exist a time window to allow for data corruption.

\textbf{Can Aurora protects its secrecy from covert channel attacks?} Even though attackers can exploit the kernel or hypervisor to control over context switches~\cite{volp_avoiding_2016}, page faults~\cite{xu_controlled-channel_2015}, page walks~\cite{DBLP:conf/uss/BulckWKPS17}, memory accesses~\cite{DBLP:conf/usenix/HahnelCP17} and LAPIC timers~\cite{VanBulck:2017:SPA:3152701.3152706} to build high-resolution covert channels, we see that our TSS events introduces random level of noise and raises the bar for covert channel attacks. Recall that Aurora allows for resets on the secure channel, enclave author may wish to reset when a monotonic counter meets a fixed upper-bound. We leave this heuristic countermeasure for users to decide.

\textbf{Can Aurora protects network stack from existing attack?} Aurora protects against general traffic analysis attack on the host side. We do not consider traffic attack from outside. Future work might put a firewall. Intrusion Detection System (IDS) or Intrusion Prevention System (IPS) into SMM layer to enhance its security. We assume cloud service provider to take common LAN security measures to protect each host, and is able to mitigate SYN flooding attacks from outside.

\subsection{Aurora's Overhead}\label{overhead}
\begin{table}
\centering
\begin{tabular}{lrrrr}
\toprule
\multirow{2}{*}{Interval} & \multicolumn{2}{c}{Time cost} & \multicolumn{2}{c}{Overhead ratio} \\\cmidrule{2-3}	 \cmidrule{4-5}
 & 78us & 152us & 78us & 152us \\
 \midrule
linux native & 8.545 & 8.635 & 0 & 0\\
1000ms & 9.045 & 9.025 & 5.85\% & 4.52\%\\
100ms & 9.605 & 9.59 & 12.40\% & 11.06\%\\
10ms & 10.18 & 10.455 & 19.13\% & 21.08\%\\
1ms & 10.16 & 10.795 & 18.90\% & 25.01\%\\
\bottomrule
\end{tabular}
\caption{Measuring Aurora's overhead by calculating Pi.}
\label{table:pi}
\end{table}

%~\cite{pi-css5}
We use pi\_css5 program to benchmark the whole system to measure the overhead introduced by Aurora. We evaluate the system slowdown by calculating the Pi number to 4,000,000 digits of precision. Since the SMM preemption time depends on the TSS response time, and the occurrence of one TSS request is unpredictable, we use two durations (78us and 152us) to emulate the real-world environments, respectively.

We use two threads to conduct this experiment. One thread (T1) is the program running Pi number calculation, the other thread (T2) is an enclave program that requests Aurora’s TSS constantly. First we calculate the Pi number without the interference oh T2 and obtained the baseline. Then we start T2 to see the slowdown of T1. We can see that when the interval decreases, the overhead also decreases. The worst case of Aurora’s overhead is around 25\%, where the TSS is requested too frequently. Such situation is rare in real world because the enclave has its own logic to process when obtaining the results from TSS. When requesting TSS at a normal frequency like 1s (the second row in the table), the overhead is no more than 6\%. We think that Aurora’s TSS is feasible in practice.

We can see from the table \ref{table:pi} that the overhead increases when the interval decreases and the duration increases. We observe that the intensive TSS requests may impact the short-life jobs. Notice that we run the benchmark on only one core of CPU. On a multiple CPU platform, we can use a dedicated CPU to provide TSS for enclave, minimizing the impact on time-sensitive tasks.

\subsection{Breakdown of Time TSS}\label{time_bench}
It is important to quantify how much time is required to attain a time TSS since the potential service latency impacts the freshness of message and indicates the response speed. For this experiment, we instrumented Aurora’s code and measured the complete time TSS for 10000 time to obtain the average value.
\begin{table}
\centering
\begin{tabular}{lr}
\toprule
Action step & Time cost(us) \\
\midrule
EPC encryption & 2\\
Copy to shared RAM & 2\\
Switch to SMM & 13\\
Copy to SMRAM & 0\\
SMRAM decryption & 3\\
Clock Service & 44\\
SMRAM encryption & 3\\
Copy to shared RAM & 0\\
Return and enter SGX & 12\\
Copy to EPC & 3\\
EPC decryption & 2\\
\bottomrule
\end{tabular}
\caption{ Breakdown of the time TSS }
\label{table:breaktime}
\end{table}

Table \ref{breaktime} shows the observed time taken for each steps, as explained in \autoref{workflow}. We can see that the clock service takes most of time. This is because the RTC driver has to read clock value twice in order to check if the clock is updating to avoid time data incorrection. The RTC involves several INS instructions to read each field of a calendar time, therefore it costs as much as 14 us to obtain a complete wall-clock time in one request. The other 4 timer costs roughly 10 us in sum. The total response time is nearly 84 us, which satisfies the requirement of real-world trusted timestamps.

\subsection{Latency of Network TSS}

\begin{figure}[t]
 \centering
\includegraphics[width=0.48\textwidth]{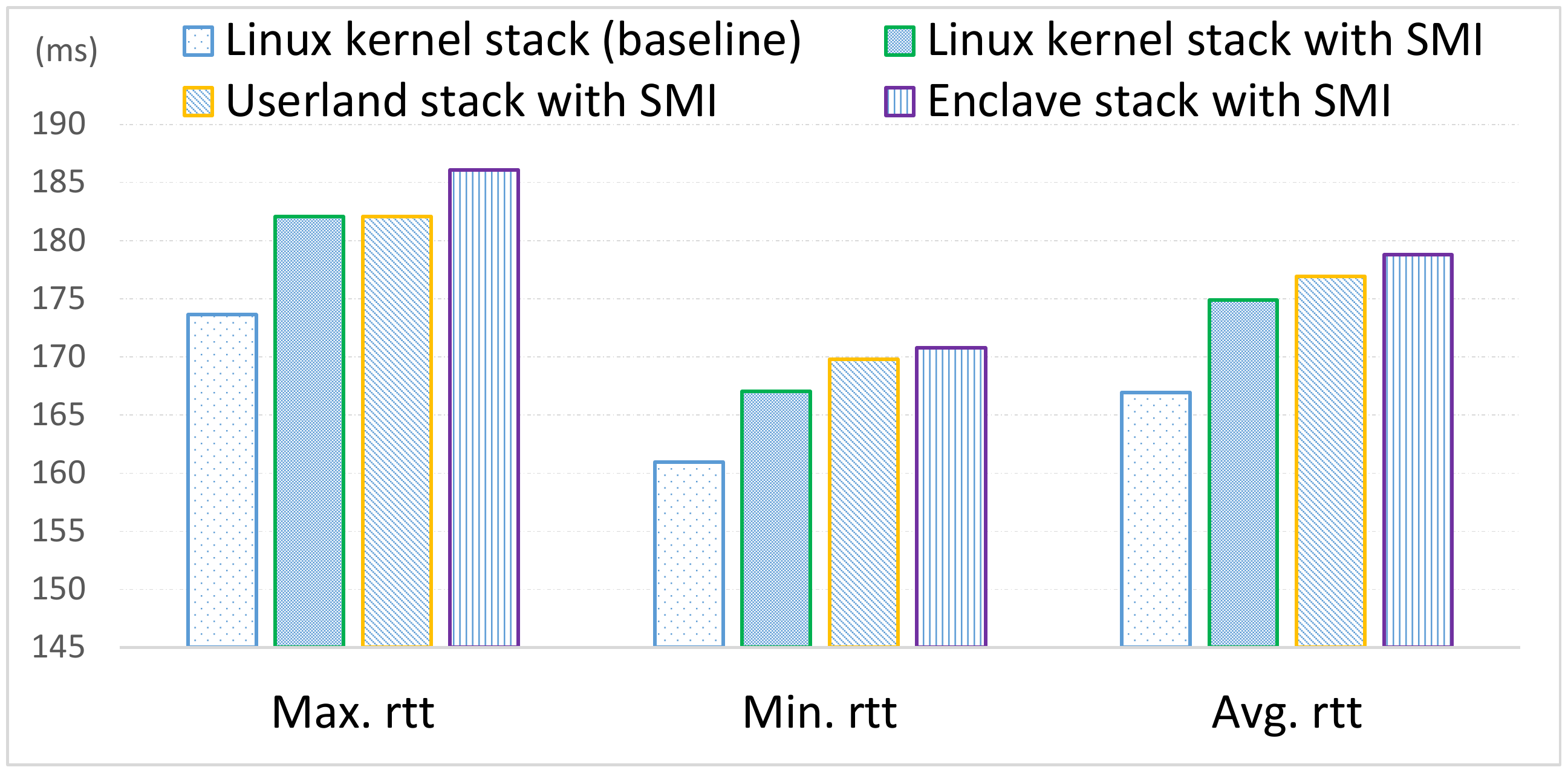} %height=0.25\textheight,width=0.78\textwidth
 \caption{Performance comparison of network stack under different circumstances.}
 \label{fig:latency}
\end{figure}

To evaluate the influence of Aurora's architectural overhead on secure network latency, we measure the enclave stack latency compared with Ubuntu 16.04 as the baseline. We study the round trip time of ICMP packets using nping. We use the command \textit{nping -icmp -c 100 -icmp-type 8 --ip-options \textbackslash x00 \textbackslash x00 \textbackslash x00 \textbackslash x00} to generate packets with special \textbf{IP\_OPTIONS} pattern. Figure \ref{fig:latency} shows the performance of network stack under different circumstances. We observe that majority of the latency is introduced by the SMM context switch, with no more than 7\% latency in total compared with the in-kernel stack.

\subsection{SQLite with Time TSS}\label{sqlite}

\begin{table*}%\small
\centering
\begin{tabular}{cccccccc}%{ |c|c|c|c|c|c|c|c| }
 \toprule
 Source & Approach & Type & Accuracy & Request Cost & Latency & Security & Related Work \\
 \midrule
 Remote Server & NTP/PTP & absolute & 1s & >100ms & high & trusted & Slick, Town Crier \\
 CSME & PSW & relative & 1s & 10.3ms & medium & trusted & SGX-Tor \\
 OS & OCALL & absolute & 1us & 6us & low & wild & Haven, Panoply \\ %\hline
 Hardware Timer & SMM & absolute & 1us & 84us & low & trusted & Aurora \\
 \bottomrule
\end{tabular}
\caption{Comparison of existing time services that can be used for enclaves. The NTP denotes Network Time Protocol, the PTP denotes Precision Time Protocol, CSME denotes Intel Converged Security and Management Engine, PSW denotes the Platform Software.}
\label{table:timeservice}
\end{table*}

A database keeps track of the activity by recoding the occurrence time for its each transaction. Without a trusted clock source, it may cost more price to maintain the event causality. SQLite is a popular database choice for local storage in application software. It is used today by several widespread browsers, operating systems, and embedded systems.

We integrated our trusted time TSS into existing SGX-SQLite~\cite{sqlite-sgx} and supports standard time APIs including \textit{time}, \textit{localtime}, \textit{utimes} and \textit{gettimeofday} for SGX-SQLite. We then eliminate its time-related \textsc{OCALLs} on the untrusted system. We evaluate the trusted time request performance by executing SQL statements \textbf{"SELECT date('now');"} and \textbf{"SELECT strftime('\%s','now');"} for 10000 times. For comparison, we also benchmark the method of using original \textsc{OCALLs} and native \textit{sgx\_get\_trusted\_time()}. Table \ref{table:timeservice} shows the overall comparison results of existing time service approaches.

Any of the approaches for time TSS can be arbitrarily delayed in that the untrusted system is in charge of the enclave thread scheduling. We treat intentionally time delay as a form of denial of service attack. Our approach outperforms remote clock and Intel’s reference clock by introducing very low latency, while achieving the same accuracy as OS time service.

\subsection{OpenSSL with Network TSS}\label{openssl}
Previous attacks~\cite{DBLP:conf/uss/SunEVLRCM15, DBLP:conf/ccs/MittalKJCB11, DBLP:conf/sp/MurdochD05} have shown that the attackers can compromise the hidden services of the Tor project by traffic analysis at the both ends of the client and server side. Preventing the network path including the protocol stacks from being traced can mitigate this attack. We integrated the OpenSSL library from the SGX-Tor~\cite{DBLP:conf/nsdi/KimHHKH17} project with our network TSS, protecting its traffic flow from observing by the untrusted systems or the privileged software.

\begin{figure}[t]
 \centering
 \includegraphics[width=0.48\textwidth]{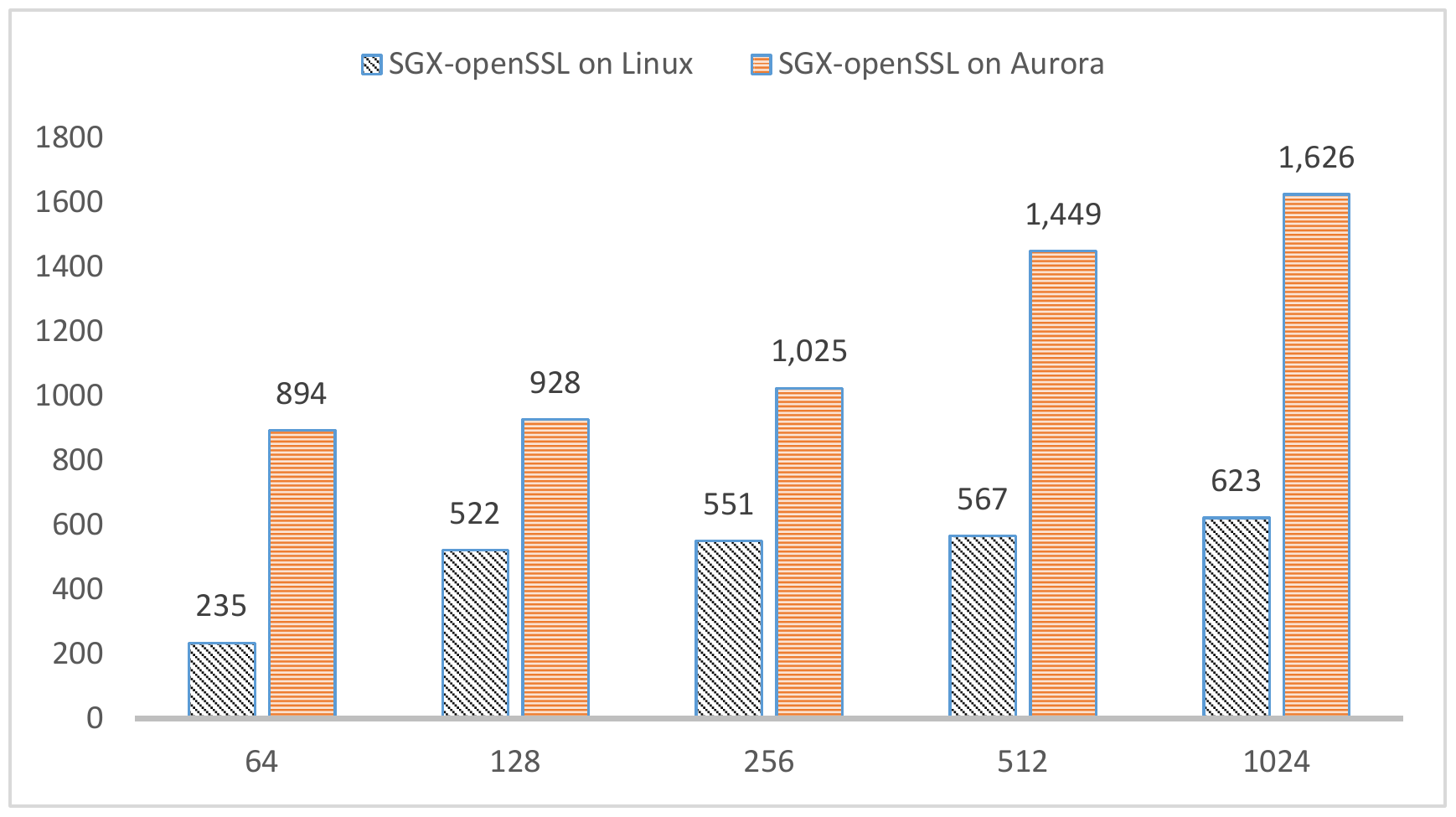} %height=0.25\textheight,width=0.78\textwidth
 \caption{Performance comparison of OpenSSL's TLS on Linux and Aurora.}
 \label{fig:openssl}
\end{figure}

We use memory pool instead of heap for lwIP to accelerate the TCP memory allocation speed. At both ends, we disable Nagle's Algorithm to maximize the throughput. Since we do not modify the SGX-OpenSSL and it uses \textit{malloc} from SGX SDK libc to process TLS messages, we set 60MB heap to avoid possible EPC paging.

We measure the overhead of to the real-world application OpenSSL. We use the Round-Trip-Time (RTT) of fixed size packets over 100 times and compute the average time for each packet’s RTT.
First, we use the native SGX-OpenSSL to establish a TLS secure session to measure the RTT time. In the implementation of native SGX-OpenSSL, it uses the OCALL to rely on untrusted socket to send the payloads. Since we focus on providing TSSes for enclaves in the cloud environment, we then replace the TLS server with Aurora-OpenSSL server, which request Aurora’s network TSS to protect its traffic from end-point observing. 
Figure \ref{fig:openssl} shows the experimental results. In our current network TSS prototype, the total time of transmit a frame is around 500us, which consists of 400us to save and restore the context (critical device control registers) and 100us to complete the transmission action itself. The reception time is approximately 200us because the driver needs traverse the RX ring buffer and check each Ethernet frame. Both of the RTT time consist of the transmission time, network driver processing time, network stack processing time and OpenSSL processing time. We ignore the propagation delay since we setup our experiment on top of the same machine and use the virtual network environments. We use the WireShark to observe the traffic flow. We observe that when the packet size exceeds 500 Bytes, the Aurora-OpenSSL introduces around 1000us. This is because the TLS introduces more packet fragmentation.

%At the time of this writing, the SGX-Tor for Linux system is not available. We plan to integrate the whole services when the Linux SGX-Tor is ready.

\section{Discussion}~\label{discussion}
\textbf{Multi-processor consideration.}
The current implementation of our prototype does not consider the multi-processor scenario. On a multi-processor platform, an SMI traps the corresponding processor into SMM mode while other processors remain in protected mode. This requires mutual exclusiveness mechanism (e.g. IPI) to synchronize I/O hardware access. Attackers may steal secrets using other processors as a covert channel.

\textbf{Verified SMM supervisor.}
Because SMM has the highest privilege, bugs and crashes in SMRAM are disastrous to the whole system. We have statically analyze our prototype using Clang Static Analyzer~\cite{clang_analyzer} and fix all vulnerabilities we found, but we cannot prove that it has no bugs that can be exploited by the adversary. We plan to formally verify our SSV in the future to make sure it will not compromise the rest of the system and regulate the SMM drivers as expected.

%\textbf{TSS access control.} SGX is fit to protect critical part of microservice containers~\cite{DBLP:journals/ieeesp/Fetzer16}. This large-scale yet decentralized model requires an MLS-like security policy enforcement~\cite{Beekman:2017:CSA:3152701.3152710}. Existing NetLabel projects such as Commercial Internet Protocol Security Option (CIPSO) can restrict end-to-end security for distributed enclave networks. We plan to extend Aurora's SSV to enforce access control lists (ACLs) on network TSS for enclaves.

\textbf{Future work.}
OpenVPN is a suitable case that will benefit from our network TSS. On the untrusted cloud, the VPN server discloses the intent of the VPN client and thereby exposes the user privacy. We plan to support network TSS for OpenVPN in the future. %We have ported OpenVPN into enclave. Despite the complex dependency on other libraries, OpenVPN cooperates with the virtual network device \textit{/dev/net/tun} supported by Linux kernel. In order to provide non-trivial system services, we need to provide proper abstractions to ease such porting.

\section{Related Work}
\textbf{Trusted I/O Path.} Zhou et al.~\cite{zhou_building_2012} uses a micro-hypervisor to build a verifiable trusted path between I/O devices and applications, assuring secrecy and authenticity on x86 systems. It leverages IOMMU to mitigate DMA attacks and the hypervisor to prevent from interrupt spoofing attacks. Our approach also defends against DMA attacks and takes over device interrupts. SGXIO~\cite{weiser_sgxio:_2017} offers generic trusted I/O paths for enclaves by combining TPM, SGX and hypervisor. Unlike SGXIO, we use the most privileged mode to build trusted paths instead of a trusted hypervisor. AuditedIO~\cite{DBLP:conf/apsys/BalakrishnanCBS17} leverages SGX and a kernel module to implement verifiable data storage for trustworthy disk I/O. However, its kernel module is not protected and may be compromised. We protect our supervisor in SMRAM and use remote attestation to ensure its authenticity. Bumpy~\cite{DBLP:conf/ndss/McCunePR09} utilizes dedicated hardware to establish trusted user input path and hence cannot generalize to other devices. TrustUI~\cite{DBLP:conf/apsys/LiMHXZCL14} splits mobile drivers into front-end and back-end facilitated by ARM TrustZone and builds a trusted path between user and Internet service. ARM TrustZone can not only isolate physical memory but also peripheral interrupts. Our solution achieves the interrupt isolation by routing desired interrupts to SMI.

\textbf{SMM-based Protection.} Our approach was inspired by Scotch~\cite{DBLP:conf/raid/LeachZW17}, which is the first work that combines SGX and SMM to monitor and audit cloud resource usage. Its motivation differs from ours. TrustLogin~\cite{zhang_trustlogin:_2015} leverages SMM for users to login a remote server without leaking sensitive credentials, even when the OS is compromised. IOCheck~\cite{DBLP:conf/esorics/ZhangWLS14} leverages an SMM monitor to check the integrity of I/O configurations and firmware at runtime, rather than providing SMM-based I/O services. Prior works like HyperCheck~\cite{DBLP:conf/raid/WangSG10} and HyperSentry~\cite{DBLP:conf/ccs/AzabNWJZS10}, Spectre~\cite{DBLP:conf/sp/ZhangLSWS15} aim to protect kernel integrity with assistance of SMM. %These works take more than 10ms to perform security check , while our solution takes 500~us at most, which introduces less overhead to the system.

\textbf{Trusted Time Service.} The SGX v1~\cite{DBLP:conf/isca/McKeenABRSSS13} disallows \textit{RDTSC} and \textit{RDTSCP} executed inside enclaves. As a compensation, Intel releases an alternative that supports trusted time and monotonic counters using Intel Management Engine. However, this service is coarse-grained and does not provide a wall-clock time. Our solution provides rather high accuracy trusted time value. Slick~\cite{DBLP:journals/corr/abs-1709-04226} achieves high-precision yet low-latency clock service by on-NIC PTP clock. However, the time source is not secure because enclaves cannot detect the tampering of NIC-Timer. By contrast, our approach can detect time attack using cross validation. Déjà Vu~\cite{DBLP:conf/ccs/ChenZRZ17} implements a reference clock thread using TSX to provide a trustworthy source of time measurement. Déjà Vu's goal is to detect interrupt-based attacks, while we provide a general time service for enclaves.

\textbf{Secure Network Service.} IX~\cite{DBLP:conf/osdi/BelayPKGKB14} offers robust protection to its network stack via privilege separation leveraging Intel VT-x technology. Arrakis~\cite{DBLP:journals/tocs/PeterLZPWKAR16} uses IOMMU to protect network device access.
SCONE~\cite{DBLP:conf/osdi/ArnautovTGKMPLM16} uses a TLS-based network shield to protect application's payloads inside containers.
SGX-Tor~\cite{DBLP:conf/nsdi/KimHHKH17} provides secure networking for Tor using in-enclave SSL/TLS. However, it depends on 57 system calls and thereby introduces large attack vectors. LightBox~\cite{DBLP:journals/corr/DuanYW17}, Trusted Click~\cite{DBLP:conf/codaspy/CoughlinKW17} and Slick~\cite{DBLP:journals/corr/abs-1709-04226} focus on securing middleboxes. By contrast, Aurora's goal is to provide end-to-end security, it protects network from L1 to L7.

\section{Conclusions}

We present Aurora, a novel architecture which can provide trusted system services for enclaves on an untrusted underlying system. Based on Aurora, we implement a time service with high-precision and attack-awareness and an end-to-end network service. To the best of our knowledge, we are the first to provide such TSSes for enclaves. Our prototype implementation demonstrates that Aurora is extensible and transparent with underlying commodity systems. Performance evaluation with real-world applications and security analysis show that Aurora is practical to provide trusted system service for SQLite and OpenSSL.

\section*{Acknowledgments}

We thank the anonymous reviewers. We are grateful to Kai Huang, Shweta Shinde for their feedback and help on Aurora.

%\section*{Availability}
%The Aurora is a free online. You can find it on the github:.
%\begin{center}
%{\tt ftp.site.dom/pub/myname/Wonderful}\\
%\end{center}

{\footnotesize \bibliographystyle{acm}
\bibliography{References}}

%\theendnotes

\end{document}